\documentclass[prx,twocolumn,preprintnumbers,amsmath,amssymb]{revtex4}

\usepackage{graphicx}% Include figure files
\usepackage{bm}% bold math
\usepackage{color}% bold math

\begin{document}

%\preprint{APS/123-QED}

\title{Nonequilibrium phonon dynamics beyond the quasiequilibrium approach}
\author{Shota Ono}
\email{shota_o@gifu-u.ac.jp}
\affiliation{Department of Electrical, Electronic and Computer Engineering, Gifu University, Gifu 501-1193, Japan}

\begin{abstract}
The description of nonequilibrium states of solids in a simplified manner is a challenge in the field of ultrafast dynamics. Here, the phonon thermalization in solids through the three-phonon scatterings is investigated by solving the Boltzmann transport equation (BTE). The numerical solution of the BTE shows that the transverse acoustic and longitudinal acoustic (LA) phonon temperatures are not well-defined during the relaxation, indicating the breakdown of the quasiequilibrium approximation. The development of hot and cold phonons and the backward energy flow from low to high energy phonons are observed in the initial and final stage of the relaxation, respectively. A minimal model is presented to relate the latter with the power-law decay of the LA phonon energy.
\end{abstract}

%\pacs{61.48.-c, 72.80.Rj, 73.20.Fz}
%61.48.-c: Structure of fullerenes and related hollow and planar molecular structures
%72.80.Rj: Fullerenes and related materials / Conductivity of specific materials
%73.20.Fz: Weak or Anderson localization / Electron states at surfaces and interfaces

\maketitle

%%%%%%%%%%%%%%%%%%%%%%%%%%%%%%%%%%%%
\section{Introduction}
%%%%%%%%%%%%%%%%%%%%%%%%%%%%%%%%%%%%
Thermalization of quasi-particles and elementary excitations in solids is a complex phenomenon because electron-electron (e-e), electron-phonon (e-ph), and phonon-phonon (ph-ph) dynamics are simultaneously involved. To develop a language for describing the nonequilibrium states in a simplified manner is highly desirable. Although the debate on this issue is still far from settled, two-temperature model (TTM) for electrons and phonons \cite{anisimov,allen} has served as a minimal model in the field of ultrafast dynamics. For example, the TTM has been widely used to study the energy relaxation of a variety of materials such as metals \cite{brorson,lin2008,brown2016,nakamura}, nanocarbons \cite{bistritzer,vilijas,ono3}, Dirac semimetals \cite{lundgren}, and warm-dense matters \cite{white2014}.

The main assumption behind the TTM is that through the e-e and ph-ph scatterings, the electrons and phonons immediately reach an equilibrium state that is characterized by the time-dependent electron and phonon temperatures, respectively \cite{allen}. However, the breakdown of the TTM in the relaxation dynamics has been addressed by several authors \cite{groeneveld,rethfeld,kabanov2008,ishida2011,mueller,baranov,ishida2016,waldecker}. This may be attributed to (i) the Pauli exclusion principle, which reduces the scattering phase space \cite{groeneveld,rethfeld,kabanov2008,mueller}, (ii) the strong electron screening, which slows the electron thermalization time \cite{mueller}, and (iii) the strong e-ph coupling, which disturbs the electron and phonon distributions significantly \cite{ishida2011,baranov}. Recently, the breakdown of the TTM due to the nonthermal phonon distribution has been reported in a layered material \cite{ishida2016} and even in aluminum \cite{waldecker}. This is because the occupation numbers of the longitudinal acoustic (LA) and transverse acoustic (TA) phonons in those solids are described by Bose-Einstein (BE) function with different temperatures, while in the TTM these are described by the same temperature. 

Recent experiments have made it possible to investigate the time-evolution of phonon distribution in solids \cite{trigo,harb}. Such achievements together with theoretical works \cite{shin,fahy} have revealed novel phonon dynamics on a picosecond time scale, such as the branch-dependent population dynamics \cite{trigo} and the phonon production by upconversion \cite{shin}.

With these nonequilibrium phonons emerging in solids, it is time to study the quasi-equilibrium treatments for phonons in detail and consider whether some intriguing rules for the phonon thermalization can be found. This is the aim of this paper. Therefore, we first present a numerical solution of the Boltzmann transport equation (BTE) for solids and discuss the phonon thermalization through the three-phonon scatterings. Then, we show that the TA and LA phonon temperatures are not well-defined during the relaxation, since the phonon distribution for each branch is not described by BE statistics. In the initial stage of the relaxation, each phonon subset develops into hot and cold phonons with time. In the final stage of the relaxation, the backward energy transfer from low to high energy regions occurs. This yields the power-law decay of the LA phonon energy, which explains the recent experimental observations \cite{ishida2016}. The relaxation behavior for each stage is illustrated by a simplified model derived from the BTE.

The rest of this paper is organized as follows. In Sec.~\ref{sec:formulation}, we formulate a theory of the phonon thermalization of solids based on the BTE considering the three-phonon scatterings. In Sec.~\ref{sec:noneq}, we examine the time-evolution of the phonon occupations, and study the approach to equilibrium. We demonstrate the breakdown of the quasi-equilibrium approximation during the relaxation, and propose a nonequilibrium function that quantitatively describes the numerical results. The relaxation dynamics in the initial and final stages are investigated by constructing simple models in Secs.~\ref{sec:initial_relaxation} and \ref{sec:final_relaxation}, respectively. In Sec.~\ref{sec:remarks}, some remarks including an interpretation of the experiment \cite{ishida2016} are presented. Finally, we summarize our conclusion in Sec.~\ref{sec:conclusion}. Technical detail concerning the fitting procedure of the numerical data is given in Appendix \ref{app:nonlinear}. Numerical simulation results with the use of the different matrix element for the three-phonon scatterings are given in Appendix \ref{sec:fconst}.

%The ph-ph scattering effect has played a crucial role in the study of phonon lifetime \cite{tamura,narasimhan,debernardi}. 
% in anharmonic monatomic lattices is investigated by solving the BTE. Numerical solution of the BTE shows 

%%%%%%%%%%%%%%%%%%%%%%%%%%%%%%%%%%%%
\section{Formulation}
\label{sec:formulation}
\begin{figure}[bbb]
\center
\includegraphics[scale=0.45,clip]{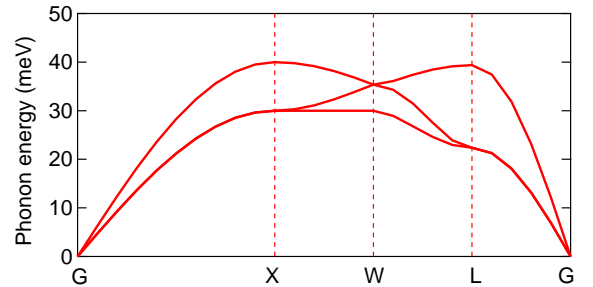}
\caption{\label{fig:phonon_band} Calculated phonon dispersion relations along the symmetry points. At X point, $\hbar\omega_{\rm X, LA}$ and $\hbar\omega_{\rm X,TA}$ are set to 40 and 30 meV, respectively.}
\end{figure}

%%%%%%%%%%%%%%%%%%%%%%%%%%%%%%%%%%%%
We study the phonon thermalization on a face-centered cubic monatomic lattice. We expand the lattice potential energy in powers of the displacement of atoms from the equilibrium position, as ${\cal V}= \sum_{p}{\cal V}_{p}$, with $p=0,2$, and 3. ${\cal V}_0$, ${\cal V}_2$, and ${\cal V}_3$ are the rigid lattice, harmonic, and anharmonic potentials, respectively. The use of ${\cal V}_{2}$ enables to compute the phonon band structure with two adjustable parameters \cite{ashcroft_mermin}, if we assume that ${\cal V}_2$ depends on the distance between nearest-neighbor atoms only. By defining the force constants $A$ and $B$ as
\begin{eqnarray}
A &=& \frac{2}{d}\frac{d{\cal V} (r)}{dr}\Big\vert_{r=d}, 
\nonumber\\
B &=& 2\left[ \frac{d^2 {\cal V} (r)}{dr^2}\Big\vert_{r=d} 
- \frac{1}{d}\frac{d{\cal V} (r)}{dr}\Big\vert_{r=d}
\right],
 \label{eq:AB}
\end{eqnarray}
where $d$ is the equilibrium nearest-neighbor distance, the phonon energies are calculated by diagonalizing the dynamical matrix given by
\begin{eqnarray}
 D = \sum_{\bm{R}} \sin^2 \left( \frac{\bm{q}\cdot \bm{R}}{2}\right) 
 \left[ A \bm{1} + B \hat{R} \hat{R} \right],
 \label{eq:dynamical_matrix}
\end{eqnarray}
where $\bm{1}$ is the $3\times3$ unit matrix and $\hat{R} \hat{R}$ is the dyadic formed from the unit vectors $\hat{R}= \bm{R}/\vert \bm{R} \vert$ with $\bm{R}$ being the nearest neighbor point vector. Given three eigenvalues $\lambda$, the phonon frequencies are given by $\omega = \sqrt{\lambda /M_i}$ with the ion mass $M_i$. The phonon energy is denoted by $\hbar\omega_{\bm{q},\mu}$, where $\hbar$ is the Planck constant, $\bm{q}$ is the wavevector, and $\mu$ is the branch index, {\it i.e.}, TA1, TA2, and LA. Note that the phonon frequency at X point is explicitly given by 
\begin{eqnarray}
\omega_{\rm X, LA} = \sqrt{\frac{8A+4B}{M_i}},
\ \ 
\omega_{\rm X, TA} = \sqrt{\frac{8A+2B}{M_i}}.
\label{eq:omegaX}
\end{eqnarray}
By setting $\hbar\omega_{\rm X, LA}=$40 meV and $\hbar\omega_{\rm X, TA}=$30 meV, which are close to the values of phonons in aluminum, the force constants $A$ and $B$ in Eq.~(\ref{eq:AB}) are, through Eq.~(\ref{eq:omegaX}), determined uniquely. Then, we obtain the phonon band structure through Eq.~(\ref{eq:dynamical_matrix}), shown in Fig.~\ref{fig:phonon_band}. 

%%%%%%%%%%%%%%%%%
%%%%%%%%%%%%%%%%%

%The whole band structure can be determined by setting the TA and LA phonon energies at X point; $\hbar\omega_{\rm X,TA}$ (doubly degenerate) and $\hbar\omega_{\rm X, LA}$.
% : each atom interacts with its 12 nearest neighbors through a pair potential
%The phonon energies are calculated by diagonalizing the dynamical matrix given by
%\begin{eqnarray}
% D = \sum_{\bm{R}} \sin^2 \left( \frac{\bm{q}\cdot \bm{R}}{2}\right) 
% \left[ A \bm{1} + B \hat{R} \hat{R} \right]
%\end{eqnarray}
%where $\bm{1}$ is the $3\times3$ unit matrix and $\hat{R} \hat{R}$ is the dyadic formed from the unit vectors $\hat{R}= \bm{R}/\vert \bm{R} \vert$ with $\bm{R}$ being the nearest neighbor point vector. $A$ and $B$ are the force constants that determine the phonon dispersion relation. Given three eigenvalues $\lambda$, the phonon frequencies are given by $\omega = \sqrt{\lambda /M_i}$ with the ion mass $M_i$. 
%Note that the phonon energy at X point is explicitly given by $\omega_{\rm LA}^{\rm X} = \sqrt{(8A+4B)/M_i}$ and $\omega_{\rm TA1(2)}^{\rm X} = \sqrt{(8A+2B)/M_i}$. By setting $\omega_{\rm LA}^{\rm X}=$40 meV and $\omega_{\rm TA1(2)}^{\rm X}=$30 meV, which are close to the values of phonons in aluminum, we obtain the phonon band structure, as shown in Fig.~\ref{fig:phonon_band}. 

The thermalization occurs through multi-phonon scatterings. The three-phonon processes are governed by the cubic term ${\cal V}_3 = \frac{1}{6}\sum A^{ijk}_{\bm{l},\bm{l}_1,\bm{l}_2} u^{i}_{\bm{l}} u^{j}_{\bm{l}_1} u^{k}_{\bm{l}_2}$, where the summation is taken over the Cartesian coordinates $i,j$, and $k$, and the lattice vectors $\bm{l},\bm{l}_1$, and $\bm{l}_2$. $u^{i}_{\bm{l}}$ is the $i$th Cartesian component of the atom displacement at the equilibrium position $\bm{l}$ and $A^{ijk}_{\bm{l},\bm{l}_1,\bm{l}_2}$ is the expansion coefficient. By introducing the phonon creation (destruction) operator $b_{\bm{q},\mu}^{\dagger}$ ($b_{\bm{q},\mu}$) for the phonon mode $(\bm{q},\mu)$, one obtains
\begin{eqnarray}
 {\cal V}_3 = \frac{1}{6}
 \sum_{\bm{q},\bm{q}_1,\bm{q}_2} \sum_{\mu,\mu_1,\mu_2}
 M_{\bm{q},\bm{q}_1,\bm{q}_2}^{\mu,\mu_1,\mu_2}
  X_{\bm{q},\mu} X_{\bm{q}_1,\mu_1} X_{\bm{q}_2,\mu_2}
 \label{eq:cubic}
\end{eqnarray}
with $X_{\bm{q},\mu} = b_{\bm{q},\mu} + b_{-\bm{q},\mu}^{\dagger}$ and the three-phonon matrix elements $M_{\bm{q},\bm{q}_1,\bm{q}_2}^{\mu,\mu_1,\mu_2}$. We apply the Fermi's golden rule for describing the probability of a transition between three-phonon states. Given no contribution from the diffusion and external field terms, the time ($t$)-evolution of the occupation number $n_{\bm{q},\mu}$ at the phonon energy $\hbar\omega_{\bm{q},\mu}$ is described by the BTE \cite{ziman,landau}
\begin{eqnarray}
 \frac{\partial n_{\bm{q},\mu}}{\partial t} 
 = \frac{2\pi}{\hbar^2N^2} 
 \sum_{\bm{q}_1\mu_1} \sum_{\bm{q}_2\mu_2}
 \left\vert M_{\bm{q},\bm{q}_1,\bm{q}_2}^{\mu,\mu_1,\mu_2} \right\vert^2
 \left(\frac{1}{2} S_{a} + S_i \right)
 \label{eq:BTE_ph}
\end{eqnarray}
with the number of unit cell $N$ and 
\begin{eqnarray}
 S_{a} &=& \left[ n_{\bm{q},\mu}^{(+)} n_{\bm{q}_1,\mu_1} n_{\bm{q}_2,\mu_2}
 - n_{\bm{q},\mu} n_{\bm{q}_1,\mu_1}^{(+)} n_{\bm{q}_2,\mu_2}^{(+)} \right]
 \nonumber\\
 &\times&
 \delta\left(\omega_{\bm{q},\mu} - \omega_{\bm{q}_1,\mu_1} -\omega_{\bm{q}_2,\mu_2} \right),
\\
  S_{i} &=& \left[ n_{\bm{q},\mu}^{(+)} n_{\bm{q}_1,\mu_1}^{(+)} n_{\bm{q}_2,\mu_2}
 - n_{\bm{q},\mu} n_{\bm{q}_1,\mu_1} n_{\bm{q}_2,\mu_2}^{(+)} \right]
 \nonumber\\
 &\times&
 \delta\left(\omega_{\bm{q},\mu} + \omega_{\bm{q}_1,\mu_1} -\omega_{\bm{q}_2,\mu_2} \right)
\end{eqnarray}
with $n_{\bm{q},\mu}^{(+)} = n_{\bm{q},\mu} + 1$. $S_{a}$ and $S_i$ denote the phonon anharmonic decay [$(\bm{q}\mu) \leftrightarrow (\bm{q}_1\mu_1) + (\bm{q}_2\mu_2)$] and inelastic scatterings [$(\bm{q}\mu) + (\bm{q}_1\mu_1) \leftrightarrow (\bm{q}_2\mu_2)$], respectively. The square of the matrix element is given by $\vert M_{\bm{q},\bm{q}_1,\bm{q}_2}^{\mu,\mu_1,\mu_2} \vert^2 = \delta_{\Delta \bm{q},\bm{G}} f(\bm{Q}, \bm{Q}_1, \bm{Q}_2)$ where $\bm{Q}=a \bm{q}/(2\pi)$ with the lattice constant $a$. $\delta_{\Delta \bm{q},\bm{G}}$ with $\Delta \bm{q} = \bm{q} \pm \bm{q}_1 - \bm{q}_2$ and the reciprocal lattice vector $\bm{G}$ indicates the crystal momentum conservation law, where $-$ for the anharmonic decay and $+$ for the inelastic scattering. The function $f$ is proportional to $\vert \tilde{A}^{ijk}_{\bm{q},\bm{q}',\bm{q}''}\vert^2 (\epsilon^{i}_{\bm{q},\mu}\epsilon^{j}_{\bm{q}_1,\mu_1}\epsilon^{k}_{\bm{q}_2,\mu_2})^2 (\omega_{\bm{q},\mu}\omega_{\bm{q}_1,\mu_1}\omega_{\bm{q}_2,\mu_2})^{-1}$, where $\tilde{A}^{ijk}_{\bm{q},\bm{q}',\bm{q}''}$ is the Fourier transformation of $A^{ijk}_{\bm{l},\bm{l}_1,\bm{l}_2}$ and $\epsilon^{i}_{\bm{q},\mu}$ is the $i$th component of the polarization vector corresponding to the mode $(\bm{q},\mu)$. Since the $(\bm{q}\mu)$-dependence of $f$ is quite complex, we simply employ the result of the continuum elasticity theory \cite{ziman,landau}; The cubic term is alternatively expressed as ${\cal V}_3 = \frac{1}{6}\sum \int B_{ijk}^{lmn} \eta_{l}^{i} \eta_{m}^{j}\eta_{n}^{k} d\bm{r}$ with $\bm{r} = (x_1,x_2,x_3)$. The summation is taken over the Cartesian coordinates $i,j,k,l,m$, and $n$. $B_{ijk}^{lmn}$ is the six rank tensor, while $\eta_{l}^{i} = \partial u^{i}(\bm{r})/\partial x_l$ serves as the second rank strain tensor, where $u^{i}(\bm{r})$ is the slowly varying displacement vector at $\bm{r}$. The Fourier transformation together with the use of the phonon creation and destruction operators yields the three-phonon Hamiltonian in a reciprocal space. The square of the matrix element is linearly proportional to $\vert \bm{q} \vert \vert \bm{q}_1 \vert \vert \bm{q}_2 \vert$ \cite{ziman,landau}. By using this expression, we define the three-phonon matrix element in Eq.~(\ref{eq:BTE_ph}) as
\begin{eqnarray}
 \vert M_{\bm{q},\bm{q}_1,\bm{q}_2}^{\mu,\mu_1,\mu_2} \vert^2 = \delta_{\Delta \bm{q},\bm{G}} w_{0}^2 \vert \bm{Q} \vert \vert \bm{Q}_1 \vert \vert \bm{Q}_2 \vert,
 \label{eq:matrix}
\end{eqnarray} 
where $w_0$ is a parameter that determines the magnitude of the matrix elements. The larger value of $w_0$ leads to the faster relaxation. Since the relaxation time in solids is usually an order of ps \cite{waldecker,trigo,harb}, we set $w_0 = 6$ meV. While the use of a realistic potential \cite{bonini,tersoff} would reveal the themalization of a specific system, such a work is beyond the scope of the present study.

The differential equation given by Eq.~(\ref{eq:BTE_ph}) is solved numerically with the time step of $\Delta t=$0.02 ps. The Dirac delta function is approximated by the Gaussian function with the broadening of 0.2 meV. The Brillouin zone is integrated with a Gamma-centered Monkhorst-Pack mesh \cite{MP} of 14$\times$14$\times$14 at each time. For the present choice of parameters, the relative error of the total energy is found to be 0.12 \% at $t=5000\Delta t$.

% Using the displacement vector $u_i (\bm{r})$ with $i=x,y,z$ at the position $\bm{r}$, the cubic term is expressed by ${\cal V}_3 = \frac{1}{6}\sum\int d\bm{r} A_{ijk}^{lmn} \left( \partial_l u_i  \right)\left( \partial_m u_j  \right)\left( \partial_n u_k  \right)$, where the summation is taken over the Cartesian coordinates $i,j,k,l,m$, and $n$. $A_{ijk}^{lmn}$ is the six rank tensor. 
 
%The $\bm{Q}$-dependence of $f$ strongly depends on the phonon model used. Instead of focusing on a specific system, we investigate the cases of $f=w_{0}^2$ and $f= w_{0}^2 \vert \bm{Q} \vert \vert \bm{Q}_1 \vert \vert \bm{Q}_2 \vert$, where $w_0$ determines the magnitude of the matrix elements. 

%Nevertheless, we confirmed that the relaxation behaviors for both cases are qualitatively the same, i.e., the presence of the power-law decay mentioned below. The results using the latter function will be demonstrated below, while those using the former are provided in the Supplemental Material \cite{suppl}.  

When a solid is excited by a pump pulse, the absorbed photon energy is transferred to the lattice via the e-ph coupling. Since the electron-LA phonon coupling is usually stronger than the electron-TA phonon coupling \cite{waldecker}, we considered the following initial condition
\begin{eqnarray}
 n_{\bm{q},\mu} = \left[ e^{\hbar\omega_{\bm{q},\mu}/(k_{\rm B}T_{\rm ini})} -1 \right]^{-1}, 
 \label{eq:initial1}
\end{eqnarray}
where $k_{\rm B}$ is the Boltzmann constant, and $T_{\rm ini}=T_{\rm low}$ for $\mu=$TA1 and TA2 and $T_{\rm ini}=T_{\rm high}$ for $\mu=$LA. Since $\hbar\omega_{\rm X,LA}=$40 meV and $\hbar\omega_{\rm X,TA}=$30 meV, we studied several initial conditions: the low temperature limit $k_{\rm B}T_{\rm low}, k_{\rm B}T_{\rm high} < \hbar\omega_{\rm X,TA}$, the high temperature limit $\hbar\omega_{\rm X,LA} < k_{\rm B}T_{\rm low}, k_{\rm B}T_{\rm high}$, and the intermediate case such as $k_{\rm B}T_{\rm low} < \hbar\omega_{\rm X,TA}, \hbar\omega_{\rm X,TA} < k_{\rm B}T_{\rm high} < \hbar\omega_{\rm X,LA}$. We also considered the Gaussian-type excitation of phonons $n_{\bm{q},\mu} = n^{(0)}(\omega_{\bm{q},\mu},T) + n^{(1)} e^{-(\hbar\omega_{\bm{q},\mu} - \epsilon)^2}$ where $n^{(0)}(\omega_{\bm{q},\mu},T)$ is the BE function with finite $T$, $\epsilon$ is the excited phonon energy that gives a peak of the distribution function, and $n^{(1)}$ is the amplitude. For example, we set $\epsilon=$ 40 meV, and $n^{(1)}=10$, assuming that the phonons with the Debye frequency are coherently excited at room $T$. Nevertheless, these initial conditions do not change the relaxation behavior qualitatively. Below, we thus set $k_{\rm B}T_{\rm low}=$ 1 meV and $k_{\rm B}T_{\rm high}=$35 meV in Eq.~(\ref{eq:initial1}).

%%%%%%%%%%%%%%%%%
\begin{figure}[t]
\center
\includegraphics[scale=0.4,clip]{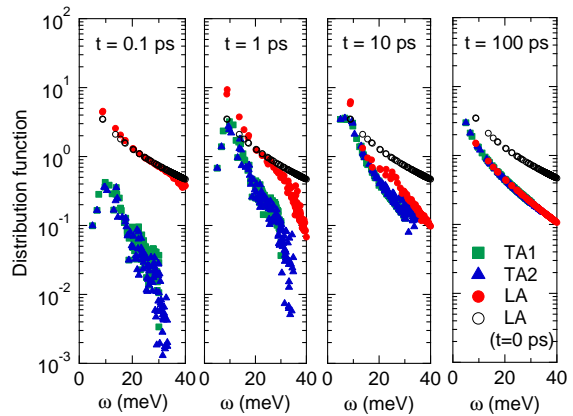}
\caption{\label{fig:time-evolution} The phonon distribution function as a function of TA1 (square), TA2 (triangle), and LA (filled circle) phonon energies for several $t$s. For comparison, the distribution function of $\mu=$LA at $t=0$ ps (open circle) is also shown.}
\end{figure}
%%%%%%%%%%%%%%%%%

%%%%%%%%%%%%%%%%%%%%%%%%%%%%%%%%%%%%
\section{Results and Discussion}
\label{sec:results}
%%%%%%%%%%%%%%%%%%%%%%%%%%%%%%%%%%%%
%%%%%%%%%%%%%%%%%%%%%%%%%%%%%%%%%%%%
\subsection{Nonequilibrium Dynamics}
\label{sec:noneq}
%%%%%%%%%%%%%%%%%%%%%%%%%%%%%%%%%%%%
Figure \ref{fig:time-evolution} shows the distribution of TA1, TA2, and LA phonon modes for $t=0.1, 1, 10$, and 100 ps. The occupation number of TA1 and TA2 phonon modes increases with time, while that of LA modes decreases. This clearly indicates that the energy is transferred from the LA to TA phonons. At $t=$100 ps, the phonon system is in equilibrium at $k_{\rm B}T=$17 meV. This is the simplest interpretation of the relaxation dynamics.

\begin{figure}[ttt]
\center
\includegraphics[scale=0.4,clip]{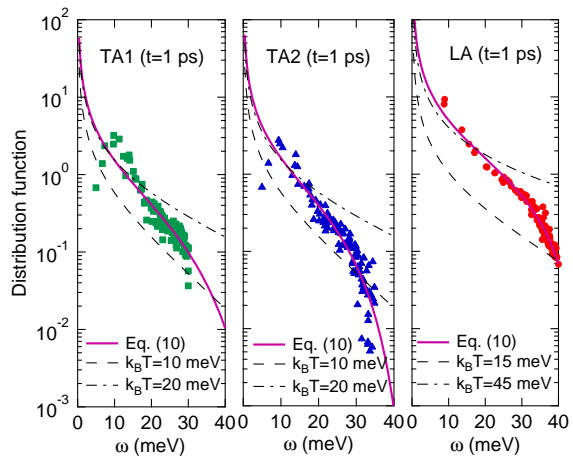}
\caption{\label{fig:1ps} Snapshot of the phonon distribution function at $t=$1 ps. The distribution function can be fit by Eq.~(\ref{eq:FDT}). The deviation from the fit is due to both the suppressed value of $f$ for $\hbar\omega\le$ 10 meV and the finite mesh size.}
\end{figure}

Figure \ref{fig:1ps} shows the phonon occupation numbers at $t=$1 ps and the BE statistics with a few lattice temperatures (dashed and dot-dashed curves). It is shown that the phonon distribution cannot be described by BE statistics, in particular in the high energy tail. This shows that the TA and LA phonon temperatures are never well-defined during the relaxation, showing the breakdown of the quasi-equilibrium treatment.

%%%%%%%%%%%%%%%%%
%%%%%%%%%%%%%%%%%

%%%%%%%%%%%%%%%%%
\begin{figure}[ttt]
\center
\includegraphics[scale=0.4,clip]{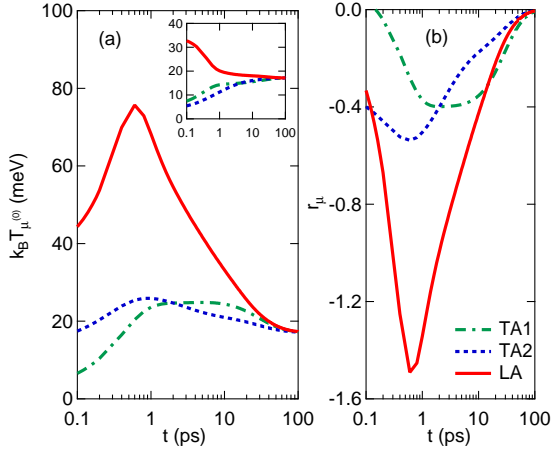}
\caption{\label{fig:kTr} The time-evolution of (a) $T_{\mu}^{(0)}$ and (b) $r_{\mu}$ ($\mu =$TA1, TA2, and LA) given in Eq.~(\ref{eq:FDT}). Inset: The $t$-dependence of $T_{\mu}^{(0)}$ with $r_{\mu}$ fixed to zero.}
\end{figure}
%%%%%%%%%%%%%%%%%

There are two reasons why the BE function fails to describe the LA and TA phonon populations. First, in Fig.~\ref{fig:time-evolution}, the population of LA phonons with $\hbar\omega_{\bm{q},{\rm LA}}\le 20$ meV and $\hbar\omega_{\bm{q},{\rm LA}} > 20$ meV, respectively, increases and decreases too much, compared to the initial distribution function, at the initial stage of relaxation. Second, the population increase of the TA phonons with $\hbar\omega_{\bm{q},{\rm TA}}\le 20$ meV is much larger than that of the TA phonons with $\hbar\omega_{\bm{q},{\rm TA}} > 20$ meV. To describe such a strong population variation with $\omega_{\bm{q},\mu}$, we consider the following function
\begin{eqnarray}
 n_{\bm{q},\mu}(t) = 
 \left[ \exp\left(\frac{\hbar\omega_{\bm{q},\mu}}
 {k_{\rm B} T_{\mu}^{(0)}(t) 
 + r_{\mu}(t) \hbar \omega_{\bm{q},\mu}}
 \right) -1 \right]^{-1}, 
 \label{eq:FDT}
\end{eqnarray}
where $T_{\mu}^{(0)}(t)$ and $r_{\mu}(t)$ are a quasi-temperature and a dimensionless parameter that characterizes the degree of the nonequilibrium of the branch $\mu$, respectively. The deviation from $r_{\mu}=0$ measures how each subset is far from equilibrium. A fit to the distribution function given by Eq.~(\ref{eq:FDT}) was performed at each time by using \texttt{minpack} \cite{minpack} (see Appendix \ref{app:nonlinear} for the numerical implemantation). As shown in Fig.~\ref{fig:1ps}, the agreement is good, indicating the validity of the form of Eq.~(\ref{eq:FDT}) to describe the nonequilibrium distribution. Note that the suppression of the population observed at $\hbar\omega_{\bm{q},\mu}\le 10$ meV in the TA phonons is due to the small $\vert M_{\bm{q},\bm{q}_1,\bm{q}_2}^{\mu,\mu_1,\mu_2} \vert^2$ for smaller $\vert\bm{q}\vert$ because it was not observed when $f=w_{0}^{2}$ is used (see also Appendix \ref{sec:fconst}).

%%%%%%%%%%%%%%%%%%%%%%%%%%%%%%%%%%%%
\subsection{Initial stage of the relaxation}
\label{sec:initial_relaxation}
%%%%%%%%%%%%%%%%%%%%%%%%%%%%%%%%%%%%
\subsubsection{Development of hot and cold phonons}
To understand the initial population dynamics, we show the $t$-dependence of $T_{\mu}^{(0)}$ and $r_{\mu}$ in Figs~\ref{fig:kTr}(a) and \ref{fig:kTr}(b), respectively. For comparison, the time evolution of $T_{\mu}^{(0)}$ with $r_{\mu}$ fixed to zero ({\it i.e.}, within the quasi-equilibrium approximation) is also shown in the inset of Fig.~\ref{fig:kTr}(a). In the latter case, $T_{\mu}^{(0)}(t)$ shows a monotonic increase and decrease for $\mu =$TA1(TA2) and LA, respectively, simply indicating that the LA phonon energy is transferred to TA phonons. In the case of $r_{\mu}(t)\ne 0$ shown in Fig.~\ref{fig:kTr}(a), $k_{\rm B}T_{\rm LA}^{(0)}(t)$ initially increases with time and takes the maximum value of 75 meV, quite higher than the initial energy, at $t\simeq 0.6$ ps. Then, $k_{\rm B}T_{\rm LA}^{(0)}(t)$ decreases slowly and approaches $17$ meV at $t=$100 ps. Conversely, $r_{\rm LA}(t)$ becomes negative and takes the minimum value of $-1.5$ at $t=0.6$ ps, after which $r_{\rm LA}(t)$ starts to approach zero shown in Fig~\ref{fig:kTr}(b). Similar behavior is observed for $\mu =$TA1 and TA2, while the variation of the two parameters as a function of $t$ is not so large, compared to the case of the LA phonon. The nonzero value of $r_\mu (t)$ indicates that each phonon subset starts to be divided into hot and cold parts until a critical $t$ ($\simeq$0.6 ps), after which they thermalize. Since $r_\mu < 0$, this can be interpreted as the development of the {\it hot} low energy phonon (LEP) and {\it cold} high energy phonon (HEP) in the initial stage of the relaxation. It would yield the backward energy flow from LEP to HEP in the final stage of the relaxation.

%%%%%%%%%%%%%%%%%
\begin{table}[bbb]
\begin{center}
\caption{Three-phonon scattering processes between LA and TA phonons}
{
\begin{tabular}{ll}\hline
%--------------------------------------------------------------------------------
$l=1$ \hspace{10mm} & $\omega_{\rm LA}' +  \omega_{\rm TA}'' \leftrightarrows  \omega_{\rm LA}$ \\ 
$l=2$  & \hspace{8.5mm} $\omega_{\rm TA}'' \leftrightarrows  \omega_{\rm LA} +  \omega_{\rm LA}'$ \\ 
$l=3$  & $\omega_{\rm TA}' +  \omega_{\rm LA}'' \leftrightarrows  \omega_{\rm LA}$ \\ 
$l=4$  & \hspace{8.5mm} $\omega_{\rm LA}'' \leftrightarrows  \omega_{\rm LA} +  \omega_{\rm TA}'$ \\ 
$l=5$  & $\omega_{\rm TA}' +  \omega_{\rm TA}'' \leftrightarrows  \omega_{\rm LA}$ \\ 
$l=6$  & \hspace{8.5mm} $\omega_{\rm TA}'' \leftrightarrows  \omega_{\rm LA}+ \omega_{\rm TA}'$ \\ 
%--------------------------------------------------------------------------------
\hline
\end{tabular}
}
\label{tab:process}
\end{center}
\end{table}
%%%%%%%%%%%%%%%%%%%%%%%%%%%%%%%%%%%%%%%%%%%%%%%%%%%

%%%%%%%%%%%%%%%%%%%%%%%%%%%%%%%%%%%%
\subsubsection{The relevant scattering processes and the upper value of the hot LEP energy}
%%%%%%%%%%%%%%%%%%%%%%%%%%%%%%%%%%%%
It is possible to determine the scattering processes relevant to the hot LEP and cold HEP creation in the LA phonon branch at $t \simeq 0$ ps. Simultaneously, the maximum of the hot LEP energy or the minimum of the cold HEP energy is also determined. To show this, we focus on the $\omega$-dependence of the LA phonon distribution function $n_{\rm LA}(\omega,t)$ and evaluate the collision term for the ph-ph scatterings at $t=0$ ps only. We then start from the BTE for $n_{\rm LA}(\omega,t)$
\begin{eqnarray}
& &\frac{\partial n_{\rm LA}(\omega,t)}{\partial t} =
\frac{\gamma_{{\rm LA}\mathchar`-{\rm TA}}^2}{\hbar^2 N^2}
\nonumber\\
&\times&
\left[
\int_{0}^{\Omega_{\rm LA}}  d\omega' D_{\rm LA} (\omega') 
+\int_{0}^{\Omega_{\rm TA}}  d\omega' D_{\rm TA} (\omega') 
\right]
\nonumber\\
&\times& 
\left[
\int_{0}^{\Omega_{\rm LA}}  d\omega'' D_{\rm LA} (\omega'') 
+\int_{0}^{\Omega_{\rm TA}}  d\omega'' D_{\rm TA} (\omega'') 
\right]
\nonumber\\
&\times& 
\left(\frac{1}{2}
{\cal F}_a + {\cal F}_i
\right),
\label{eq:dist_LA}
\end{eqnarray}
with the phonon density-of-states (DOS) $D_{\mu} (\omega)$ $(\mu=$ LA or TA), the averaged three-phonon Hamiltonian matrix elements $\gamma_{{\rm LA}\mathchar`-{\rm TA}}$ between the LA and TA phonons, and
\begin{eqnarray}
{\cal F}_a 
&=&  \Big[n_{\rm LA}^{+}(\omega) n_{\mu'}(\omega')n_{\mu''}(\omega'') 
\nonumber\\
&-& n_{\rm LA}(\omega) n_{\mu'}^{+}(\omega') n_{\mu''}^{+}(\omega'')\Big] 
\delta (\omega - \omega' - \omega''),
\nonumber\\
{\cal F}_i &=&  \Big[n_{\rm LA}^{+}(\omega) n_{\mu'}^{+}(\omega')n_{\mu''}(\omega'') 
\nonumber\\
&-& n_{\rm LA}(\omega) n_{\mu'}(\omega') n_{\mu''}^{+}(\omega'')\Big]
\delta (\omega + \omega' - \omega'')
\label{eq:F_ai}
\end{eqnarray}
with $n_{\mu}^{+}(\omega) = n_{\mu}(\omega) + 1$. $\mu'$ and $\mu''$ in Eq.~(\ref{eq:F_ai}) are the mode index corresponding to $\omega'$ and $\omega''$, respectively: For example, when the phonon mode with the frequency $\omega'$ is the TA mode, $\mu' =$TA. Since we focus on the relaxation at $t= 0$ ps, $n_{\rm LA}(\omega), n_{\mu'}(\omega')$, and $n_{\mu''}(\omega'')$ in the right hand side (r.h.s.) in Eq.~(\ref{eq:F_ai}) can be approximated by the initial distribution function, {\it i.e.}, the Bose distribution function with the temperature $T_{\rm ini} = T_{\rm low}$ for TA modes and $T_{\rm ini} = T_{\rm high}$ for LA modes [see Eq.~(\ref{eq:initial1})]. If $\mu' = \mu'' =$LA, no scatterings contribute to the collision term in Eq.~(\ref{eq:dist_LA}). This is because all the distribution functions $n_{\rm LA}(\omega), n_{\mu'}(\omega')$, and $n_{\mu''}(\omega'')$ are associated with the same temperature. Then, the r.h.s. in Eq.~(\ref{eq:dist_LA}) are decomposed into six terms $P(l)$ with $l=1,2,3,4,5$, and 6. Table \ref{tab:process} lists the scattering processes that contributes to the r.h.s. in Eq.~(\ref{eq:dist_LA}). They are explicitly given as
\begin{eqnarray}
P(1) &=& 
\int_{0}^{\Omega_{\rm LA}}  d\omega' D_{\rm LA} (\omega') 
\int_{0}^{\Omega_{\rm TA}}  d\omega'' D_{\rm TA} (\omega'') 
\frac{{\cal F}_a }{2},
\nonumber\\
P(2) &=& 
\int_{0}^{\Omega_{\rm LA}}  d\omega' D_{\rm LA} (\omega') 
\int_{0}^{\Omega_{\rm TA}}  d\omega'' D_{\rm TA} (\omega'') 
{\cal F}_i,
\nonumber\\
P(3) &=& P(1),
\nonumber\\
P(4) &=& 
\int_{0}^{\Omega_{\rm TA}}  d\omega' D_{\rm TA} (\omega') 
\int_{0}^{\Omega_{\rm LA}}  d\omega'' D_{\rm LA} (\omega'') 
{\cal F}_i,
\nonumber\\
P(5) &=& 
\int_{0}^{\Omega_{\rm TA}}  d\omega' D_{\rm TA} (\omega') 
\int_{0}^{\Omega_{\rm TA}}  d\omega'' D_{\rm TA} (\omega'') 
\frac{{\cal F}_a}{2},
\nonumber\\
P(6) &=& 
\int_{0}^{\Omega_{\rm TA}}  d\omega' D_{\rm TA} (\omega') 
\int_{0}^{\Omega_{\rm TA}}  d\omega'' D_{\rm TA} (\omega'') 
{\cal F}_i.
\nonumber\\
\label{eq:scatt}
\end{eqnarray}
Here $P(l)$ with odd and even $l$ indicates the anharmonic decay and the inelastic scattering of the phonon mode with $\omega$, respectively. We apply the Debye model for the LA and TA phonons, where the phonon DOS for the LA and TA phonons are given by 
\begin{eqnarray}
 D_{\rm LA} (\omega)&=&
  \frac{3N\omega^2}{\Omega_{\rm LA}^3}\theta_{H}(\Omega_{\rm LA}- \omega),
  \nonumber\\
  D_{\rm TA} (\omega)&=&
  \frac{6N\omega^2}{\Omega_{\rm TA}^3}\theta_{H}(\Omega_{\rm TA} - \omega),
  \label{eq:Debye}
\end{eqnarray}
respectively, with the Heaviside step function $\theta_{H}(\omega)$. As shown in Fig.~\ref{fig:phonon_band}, $\hbar \Omega_{\rm LA}$ and $\hbar \Omega_{\rm TA}$ are set to be 40 and 30 meV, respectively.

%%%%%%%%%%%%%%%%%
\begin{figure}[hhh]
\center
\includegraphics[scale=0.4,clip]{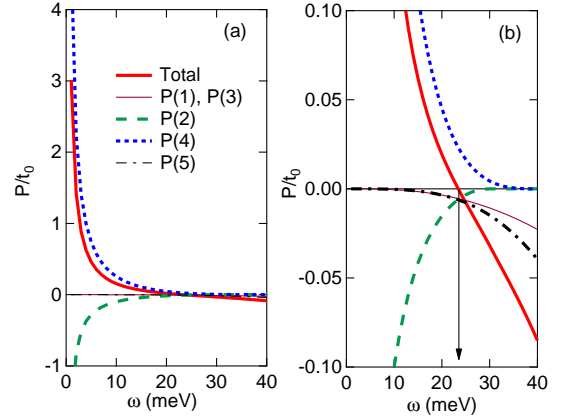}
\caption{\label{fig:collision} (a) The $\omega$-dependence of $P_{\rm tot}$ and $P(i)$ for $i=1,2,3,4$, and 5. (b) The magnified view of (a) for small $P_{\rm tot}$ and $P(i)$. $t_0$ is set to 1 ps.}
\end{figure}
%%%%%%%%%%%%%%%%%

Figure \ref{fig:collision}(a) shows the $\omega$-dependence of $P(i)$ with $i=$1, 2, 3, 4, 5, and the sum of the contribution $P_{\rm tot}\left[=\sum_{i=1}^{6}P(i)\right]$. $P(1)$ is exactly the same as $P(3)$. $P(6)$ is not shown because it is negligibly small. For smaller $\omega$, $P(2)$ and $P(4)$, related to the inelastic scattering processes, are negative and positive, respectively, so that they are cancelled out partly. This yields the positive value of $P_{\rm tot}$ for smaller $\omega$, indicating the creation of the hot LEP for $t>0$. A magnified view [Figure \ref{fig:collision}(b)] shows that $\omega$-$P_{\rm tot}$ curve crosses zero at $\hbar\omega = \hbar\Delta_{\rm LA} \simeq 23$ meV (black arrow) because of the negative values of $P(1)$, $P(3)$, and $P(5)$ that originate from the anharmonic decay processes. This, in turn, indicates the creation of the cold HEP above $\omega = \Delta_{\rm LA}$. Notice that the value of $\hbar\Delta_{\rm LA}$ is almost the same as the upper value of the hot LEP energy, shown in  Figs.~\ref{fig:time-evolution} and \ref{fig:1ps}.

%We can show that the first and second terms in Eq.~(\ref{eq:LA_TA}) are large positive for smaller $\omega$ and large negative for larger $\omega$, respectively. Each term is further decomposed into three terms. 
%All contributions from the six terms are provided in the Supplemental Material \cite{suppl}. As a result, the sign of the r.h.s. in Eq.~(\ref{eq:LA_TA}) changes across a critical $\hbar\omega = \hbar\Delta_{\rm LA}\simeq$23 meV (i.e., the cutoff energy of the LEP), yielding the creation of the hot LEP and cold HEP.

%%%%%%%%%%%%%%%%%
\begin{figure}[ttt]
\center
\includegraphics[scale=0.45,clip]{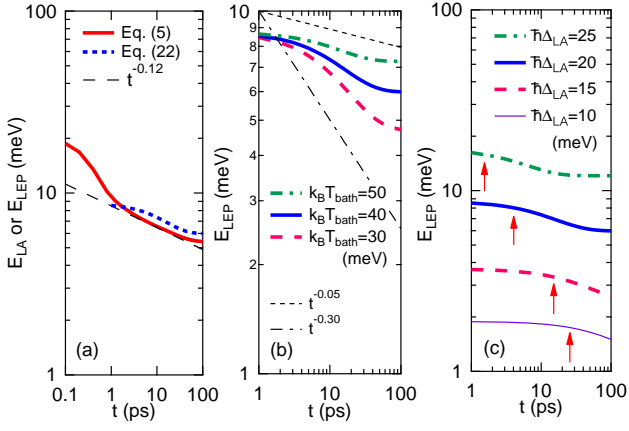}
\caption{\label{fig:power} (a) The time-evolution of the LA phonon energy $E_{\rm LA}(t)$ and the LEP energy $E_{\rm LEP}(t)$ calculated by using Eqs.~(\ref{eq:BTE_ph}) and (\ref{eq:theta}), respectively. The curves proportional to $t^{-\alpha}$ with $\alpha=$0.12 is also shown. The $t$-dependence of $E_{\rm LEP}$ for various (b) $k_{\rm B}T_{\rm bath}$ and (c) $\hbar\Delta_{\rm LA}$.}
%  for the initial conditions $(k_{\rm B}T_l, k_{\rm B}T_h)=$(1,35) meV
\end{figure}
%%%%%%%%%%%%%%%%%

%%%%%%%%%%%%%%%%%%%%%%%%%%%%%%%%%%
\subsection{Final stage of the relaxation}
\label{sec:final_relaxation}
%%%%%%%%%%%%%%%%%%%%%%%%%%%%%%%%%%
%%%%%%%%%%%%%%%%%%%%%%%%%%%%%%%%%%
\subsubsection{Power-law decay and Backward energy flow}
%%%%%%%%%%%%%%%%%%%%%%%%%%%%%%%%%%
Figure \ref{fig:power}(a) shows the time-evolution of the total LA phonon energy per a unit cell $E_{\rm LA}(t) = \sum_{\bm{q}} \hbar\omega_{\bm{q},{\rm LA}} n_{\bm{q},{\rm LA}}(t)/N$. The magnitude of $E_{\rm LA}$ decreases with time and converges to the value of 5.4 meV at $t=$100 ps. Interestingly, the power-law behavior is observed from $t=2$ to $45$ ps; $E_{\rm LA}(t) \propto t^{-p}$ with $p=0.12$. Below, we show a model to relate the power-law relaxation with the backward energy flow from the hot LEPs to cold HEPs just before reaching the equilibrium. Note that the phonon dynamics in lattices with a basis would also be described by the present model because most of the optical phonons must have decayed into acoustic phonons in the final relaxation. 

We assume that the nonequilibrium distribution function for the LA phonons is expressed as \cite{SCs}
\begin{eqnarray}
 n_{\bm{q},{\rm LA}} &=& 
 \left[ e^{\hbar\omega_{\bm{q},{\rm LA}}/(k_{\rm B}T^{*})} -1 \right]^{-1},
 \label{eq:noneq_dist}
\end{eqnarray}
with the effective temperature $T^*=\theta_{\rm LA}(t)$ for $\omega_{\bm{q},{\rm LA}} \le \Delta_{\rm LA}$ ({\it i.e.}, LEP) and $T^{*}=T_{\rm bath}$ for $ \omega_{\bm{q},{\rm LA}} > \Delta_{\rm LA}$ ({\it i.e.}, HEP). $T_{\rm bath}$ is time-independent satisfying $\theta_{\rm LA}(t) > T_{\rm bath}$, that is, the HEPs serve as a thermal bath. 

\begin{figure}[bbb]
\center
\includegraphics[scale=0.35,clip]{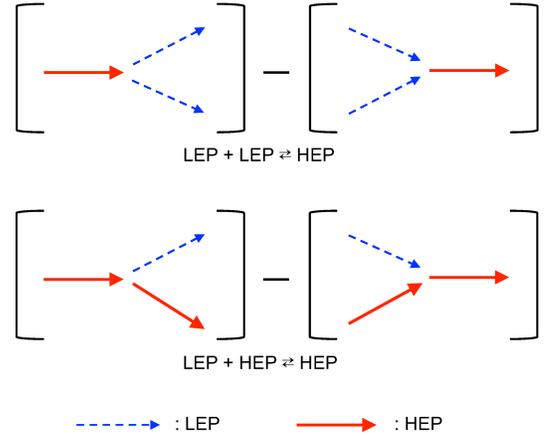}
\caption{\label{fig:phph} Schematic illustration of the inelastic scattering process expressed by ${\cal S}_i$ in Eq.~(\ref{eq:ELA-t3}). These contribute to the collision integral given by the r.h.s. of Eq.~(\ref{eq:ELA-t2}).}
\end{figure}

The time-evolution of the total energy of the LEP is given by
\begin{eqnarray}
\frac{\partial E_{\rm LEP}(t)}{\partial t} =
\frac{1}{N}
\sum_{\bm{q}}{}^{'} 
\hbar \omega_{\bm{q},{\rm LA}}
\frac{\partial n_{\bm{q},{\rm LA}}}{\partial t},
\label{eq:ELA-t}
\end{eqnarray}
where the summation is taken over all the wavevectors satisfying $\hbar \omega_{\bm{q},{\rm LA}} \le \hbar\Delta_{\rm LA}$. By substituting Eq.~(\ref{eq:BTE_ph}) into Eq.~(\ref{eq:ELA-t}) and transforming the summation into the integrals with respect to $\omega$, one find 
\begin{eqnarray}
& &\frac{\partial E_{\rm LEP}(t)}{\partial t} 
=
\frac{\gamma_{{\rm LEP}\mathchar`-{\rm HEP}}^{2}}{\hbar^2 N^3}
\int_{0}^{\Delta_{\rm LA}} d\omega D_{\rm LA} (\omega) \hbar\omega
\nonumber\\
&\times&
\left[
\left(\int_{0}^{\Delta_{\rm LA}} + \int_{\Delta_{\rm LA}}^{\Omega_{\rm LA}}\right)
 d\omega' D_{\rm LA} (\omega') 
\right]
\nonumber\\
&\times&
\left[
\left(\int_{0}^{\Delta_{\rm LA}} + \int_{\Delta_{\rm LA}}^{\Omega_{\rm LA}}\right)
 d\omega'' D_{\rm LA} (\omega'') 
\right]
\nonumber\\
&\times&
\left(\frac{1}{2}
{\cal S}_a 
+ {\cal S}_i 
\right),
\label{eq:ELA-t2}
\end{eqnarray}
with the phonon DOS $D_{\rm LA} (\omega)$ given by Eq.~(\ref{eq:Debye}). $\gamma_{{\rm LEP}\mathchar`-{\rm HEP}}$ is the three-phonon matrix element between the LEP and HEP, and is approximated to a constant value because the presence of the LEP is restricted to a relatively small region of the first Brillouin zone. ${\cal S}_a $ and ${\cal S}_i$ are the collision terms for the anharmonic decay and the inelastic scattering, respectively, and are explicitly given as 
\begin{eqnarray}
{\cal S}_a 
&=&  \Big[n_{\rm LA}^{+}(\omega) n_{\rm LA}(\omega')n_{\rm LA}(\omega'')
\nonumber\\
&-& n_{\rm LA}(\omega) n_{\rm LA}^{+}(\omega') n_{\rm LA}^{+}(\omega'') \Big] 
 \delta (\omega - \omega' - \omega'')
\nonumber\\
{\cal S}_i &=&  \Big[n_{\rm LA}^{+}(\omega) n_{\rm LA}^{+}(\omega') n_{\rm LA}(\omega'') 
\nonumber\\
&-& n_{\rm LA}(\omega) n_{\rm LA}(\omega') n_{\rm LA}^{+}(\omega'') \Big]
\delta (\omega + \omega' - \omega'').
\nonumber\\
\label{eq:ELA-t3}
\end{eqnarray}
Due to the energy conservation law, we may consider the inelastic scattering term ${\cal S}_i$ in Eq.~(\ref{eq:ELA-t3}) only. Furthermore, the three-phonon scattering process contributes to the collision term ${\cal S}_i$ only when one of the phonon states is different from the others: LEP$+$LEP$\leftrightarrows$HEP and LEP$+$HEP$\leftrightarrows$HEP (see Fig.~\ref{fig:phph}). As we will show below, only the former process is relevant to the appearance of the power-law decay. Thus we discarded the latter process to construct a minimal model. 

%%%%%%%%%%%%%%%%%
%%%%%%%%%%%%%%%%%

To derive the rate equation for $\theta_{\rm LA}(t)$ by considering the process of LEP$+$LEP$\leftrightarrows$HEP, we use the approximation $n_{\rm LA}(\omega) \simeq k_{\rm B}\theta_{\rm LA}/(\hbar\omega)$ and $k_{\rm B}T_{\rm bath}/(\hbar\omega)$ depending on the magnitude of $\omega$. Then, the collision term is given by 
\begin{eqnarray}
& &\frac{31G}{66}(k_{\rm B}T_{\rm bath})(\hbar \Delta_{\rm LA})
+ G(k_{\rm B}T_{\rm bath})(k_{\rm B} \theta_{\rm LA})
\nonumber\\
&-& G(k_{\rm B} \theta_{\rm LA})^2
\label{eq:coll1}
\end{eqnarray}
with
\begin{eqnarray}
 G = \frac{297 \gamma_{{\rm LEP}\mathchar`-{\rm HEP}}^{2}
 \Delta_{\rm LA}^{7}}{40 \hbar^3\Omega_{\rm LA}^9}.
\end{eqnarray}
Using Eq.~(\ref{eq:coll1}) and the expression of the LEP energy
\begin{eqnarray}
 E_{\rm LEP} = (k_{\rm B}\theta_{\rm LA})
 \left( \frac{\Delta_{\rm LA}}{\Omega_{\rm LA}}\right)^3,
\end{eqnarray}
one obtains the rate equation for $\theta(\tau) = \theta_{\rm LA}(t)/\theta_0$ with the dimensionless time $\tau = t / t_0$
\begin{equation}
\frac{\partial \theta (\tau)}{\partial \tau} 
 = a + b\theta(\tau) - c \left[\theta(\tau)\right]^2,
 \label{eq:theta}
\end{equation}
where 
\begin{eqnarray}
 \frac{a}{d} &=& \frac{31}{66} \left(\frac{T_{\rm bath}}{\theta_0}\right) \left(\frac{\hbar\Delta_{\rm LA}}{k_{\rm B}\theta_0}\right),  \  %+ \frac{2}{3}\left(\frac{T_{\rm bath}}{\theta_0}\right)^2
  \frac{b}{d} = \frac{T_{\rm bath}}{\theta_0}, \ 
  \frac{c}{d} = 1
    \nonumber\\
  d &=& \frac{297 \gamma_{{\rm LEP}\mathchar`-{\rm HEP}}^{2}(k_{\rm B}\theta_0)}{40 (\hbar\Omega_{\rm LA})^3}
  \left(\frac{\Delta_{\rm LA}}{\Omega_{\rm LA}}\right)^3 \left(\Delta_{\rm LA} t_0\right).
\end{eqnarray}
Negative sign of the third term in Eq.~(\ref{eq:theta}) leads to the decrease in the LEP temperature due to the process of ${\rm LEP} + {\rm LEP} \rightleftarrows {\rm HEP}$. The analytical solution of Eq.~(\ref{eq:theta}) is expressed as
\begin{eqnarray}
\theta(\tau) = \theta_{\infty} 
\left[
\frac{1+g e^{(b-2c\theta_\infty)\tau}}
{1-\cfrac{g \theta_\infty}{\theta_\infty - b/c}e^{(b-2c\theta_\infty)\tau}}
\right]
 \label{eq:sol}
\end{eqnarray}
with
\begin{eqnarray}
\theta_{\infty} = \frac{b+\sqrt{b^2 + 4ac}}{2c}
\end{eqnarray}
and 
\begin{eqnarray}
g = \cfrac{\theta(0) - \theta_{\infty}}
{\theta_\infty \left[1 + \cfrac{\theta(0)}{\theta_\infty - b/c}\right]}.
\end{eqnarray}
Since $b-2c\theta_\infty<0$ in Eq.~(\ref{eq:sol}), $\theta (\tau) \rightarrow \theta_{\infty}$ in the limit of $\tau \rightarrow \infty$. 

Before showing the comparison between the analytical and numerical results, we show that the process of LEP$+$HEP$\leftrightarrows$HEP does not cause the power-law decay. By performing a similar calculation above, the collision term is given by
\begin{eqnarray}
& &\frac{31G}{66}(k_{\rm B}T_{\rm bath})(\hbar \Delta_{\rm LA})
+ \frac{2G}{3}(k_{\rm B}T_{\rm bath})^2
\nonumber\\
&-& \frac{13G}{33} (k_{\rm B}T_{\rm bath}) (k_{\rm B} \theta_{\rm LA}).
\label{eq:coll2}
\end{eqnarray}
If there is a contribution from LEP$+$HEP$\leftrightarrows$HEP only, the rate equation for $\Theta(\tau)=\theta_{\rm LA}(t)/\theta_0$ is written as
\begin{equation}
\frac{\partial \Theta (\tau)}{\partial \tau} 
 = \alpha - \beta\Theta(\tau),
\label{eq:rate2} 
\end{equation}
where $\alpha$ and $\beta$ are positive values that depend on $T_{\rm bath}$ and $\Delta_{\rm LA}$. The analytical solution is simply given by 
\begin{equation}
\Theta(\tau) = \Theta_{\infty} (1- e^{-\beta \tau}) + \Theta (0) e^{-\beta \tau}
\end{equation}
with $\Theta_{\infty} = \alpha/\beta$. No choices of $\alpha$ and $\beta$ yield the power-law decay observed in Figs.~\ref{fig:power} and \ref{fig:power_app} (below). 

%%%%%%%%%%%%%%%%%%%%%%%%%%%%%%%%%%
\subsubsection{Comparison with numerical simulations}
%%%%%%%%%%%%%%%%%%%%%%%%%%%%%%%%%%
We set $t_0=$1 ps, $\hbar\Omega_{\rm LA}=$40, $\gamma_{{\rm LEP}\mathchar`-{\rm HEP}}=$1.5, $k_{\rm B}T_{\rm bath} = 40$, $\hbar \Delta_{\rm LA} = 20$, and the initial temperature $k_{\rm B}\theta(0)=$70 in units of $k_{\rm B}\theta_0=$1 meV. Then, the time-evolution of $E_{\rm LEP}(t)$ calculated from Eq.~(\ref{eq:theta}) is in agreement with that of $E_{\rm LA}(t)$ calculated from Eq.~(\ref{eq:BTE_ph}), as shown in Fig.~\ref{fig:power}(a). This clearly indicates that the power-law decay can be understood as the backward energy transfer from hot LEPs to cold HEPs. The value of the exponent $p$ is determined by $k_{\rm B}T_{\rm bath}$. In fact, $p$ decreases from 0.3 to 0.05 when $k_{\rm B}T_{\rm bath}$ is increased from 30 to 50 in units of $k_{\rm B}\theta_0$, as shown in Fig.~\ref{fig:power}(b). This is because the ratio of $E_{\rm LEP}(t=1{\rm ps})/E_{\rm LEP}(t=100 {\rm ps})$ is smaller for larger $T_{\rm bath}$. The onset of the power-law decay (arrows) is delayed with decreasing $\hbar\Delta_{\rm LA}$ because the energy exchange rate between the LEP and HEP is suppressed, as shown in Fig.~\ref{fig:power}(c). 

%This would explain the exponent difference between the case of $f= w_{0}^2 \vert \bm{Q} \vert \vert \bm{Q}_1 \vert \vert \bm{Q}_2 \vert$ ($p=0.12$) and $f=w_{0}^2$ ($p=0.15$) shown in Fig.~S5.

%(see Fig.~\ref{fig:phph_1})
%%%%%%%%%%%%%%%%%
%\begin{figure}[ttt]
%\center
%\includegraphics[scale=0.25,clip]{Fig_5.eps}
%\caption{\label{fig:phph_1} The equation for describing $\partial E_{\rm LEP}/\partial t$. The collisions involving two LEP and one HEP are relevant to the appearance of the power-law decay.}
%\end{figure}
%%%%%%%%%%%%%%%%%

%%%%%%%%%%%%%%%%%
\begin{figure}[ttt]
\center
\includegraphics[scale=0.35,clip]{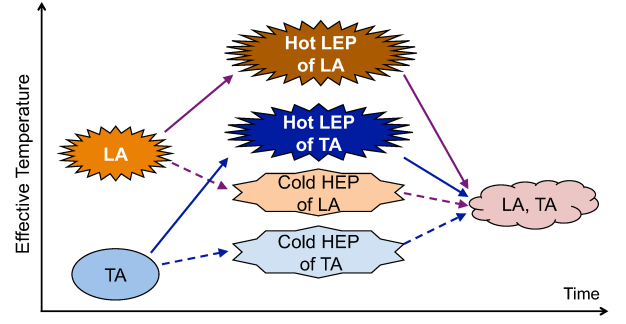}
\caption{\label{fig:models} Schematic illustration of the phonon thermalization. The direction of the energy flow is reversed in the long-$t$ limit, which leads to the power-law decay shown in Fig.~\ref{fig:power}.}
\end{figure}
%%%%%%%%%%%%%%%%%

% while the slight difference may be attributed to the presence of the e-ph coupling or the band structure effect. If the initial temperature of the LA and TA phonons is higher than the Debye temperature ({\it i.e.,} $\hbar\omega_{{\rm X, LA}}/k_{\rm B}$ in this case), the value of $p$ is very small: The result for the case of $k_{\rm B}T_l=60$ meV and $k_{\rm B}T_h=80$ meV is shown in Fig.~\ref{fig:power}(b). Thus, both the low-temperature and long-time limits are necessary to observe the power law decay. Although the emergence of the power-law relaxation may be related to the backward energy transfer from low to high energy region, which may be a rare event, there is a need of developing a theory for determining the magnitude of $p$.

%%%%%%%%%%%%%%%%%%%%%%%%%%%%%%%%%%
\subsection{Some remarks}
\label{sec:remarks}
%%%%%%%%%%%%%%%%%%%%%%%%%%%%%%%%%%
As demonstrated in Sec.~\ref{sec:final_relaxation}, the total LA phonon energy shows a power-law decay in the final stage of the relaxation. This would give an interpretation of the recent experiment by Ishida {\it et al} \cite{ishida2016}. They have studied the electron relaxation dynamics of SrMnBi$_2$ by using the time-resolved angle-resolved photoemission spectroscopy. A power-law decay of the electron energy has been observed in the final stage of the relaxation, while the TTM-like behavior has been observed in the initial stage of the relaxation. By assuming the presence of the phonon-bottleneck effect \cite{kabanov1999,rothwarf}, where the electron relaxation is regarded as the LA phonon relaxation, the power-law decay observed in the experiment can be interpreted as a backward energy flow from the hot LEP to the cold HEP. We hope that the exponent variation, as shown in Figs.~\ref{fig:power}(b) and \ref{fig:power}(c), is observed in future experiments.

We can visualize the phonon relaxation for each branch, if we consider $T_{\mu}^{(0)}(t) + r_{\mu}(t) \hbar \omega_{\bm{q},\mu}/k_{\rm B}$ in Eq.~(\ref{eq:FDT}) as an effective temperature. The combined use of Eq.~(\ref{eq:FDT}) and Fig.~\ref{fig:kTr} reveals the phonon development into hot LEP and cold HEP, followed by the thermalization involving the backward energy transfer from the former to the latter, as shown in Fig.~\ref{fig:models}. On the other hand, a monotonic evolution is only revealed with the use of $r_{\mu}(t) = 0$, as shown in the inset of Fig.~\ref{fig:kTr}(a).

%Furthermore, $\bm{q}$-dependent phonon dynamics could be investigated, whose dynamics would be compared with the nonequilibrium atomic vibrations obtained by the x-ray scattering experiments \cite{trigo}. 

Based on the TTM, Brorson {\it et al}. determined the e-ph coupling of several superconductors except aluminum from femtosecond time-resolved experiments \cite{brorson}. Recently, Waldecker {\it et al}. proposed a nonthermal lattice model (NLM) to study the energy flow in photoexcited aluminum beyond the TTM \cite{waldecker}. In the model, the phonon distribution is still expressed as a sum of thermal distributions of the three acoustic phonon branches, equivalent to the quasi-equilibrium approach. They demonstrated that the determination of the e-ph coupling from time-resolved experiments by means of the NLM leads to sufficiently correct values. However, the numerical solution presented in this paper clearly shows that each phonon subset is not in thermal equilibrium during the relaxation. In this way, our results pose a fundamental question why the TTM and the NLM could serve as a good model for quantitatively determining the e-ph coupling of metals \cite{brorson,waldecker}. This would be an open question.

%%%%%%%%%%%%%%%%%%%%%%%%%%%%%%%%%%
\section{Conclusion}
\label{sec:conclusion}
%%%%%%%%%%%%%%%%%%%%%%%%%%%%%%%%%%
In conclusion, through numerical simulations, we have demonstrated the breakdown of the quasi-equilibrium approach during the phonon thermalization. The analyses have revealed the phonon development into two subgroups and the backward energy flow between them in the initial and the final stage of the relaxation, respectively. The latter yields the power-law decay of the LA phonon energy, which explains the recent experimental observations \cite{ishida2016}. The present study could be a crucial ingredient to construct a model beyond TTM. Our model can be generalized to incorporate the several effects; the e-e and e-ph scatterings and more realistic situations such as the presence of the optical phonon modes and the optical excitations. This will be a future work.

\begin{acknowledgments}
The author would like to thank Y. Ishida for many enlightening discussions. This study is supported by a Grant-in-Aid for Young Scientists B (No. 15K17435) from JSPS.
\end{acknowledgments}

\appendix

%%%%%%%%%%%%%%%%%
\begin{figure}[ttt]
\center
\includegraphics[scale=0.4,clip]{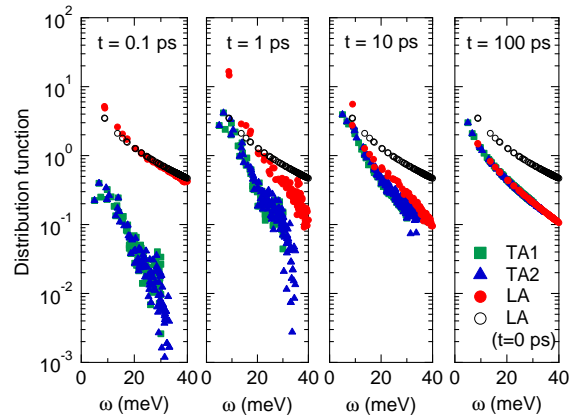}
\caption{\label{fig:time-evolution_app} Same as Fig.~\ref{fig:time-evolution} but for the case of $f=w_{0}^{2}$.}
\end{figure}
%%%%%%%%%%%%%%%%%
%%%%%%%%%%%%%%%%%
\begin{figure}[ttt]
\center
\includegraphics[scale=0.4,clip]{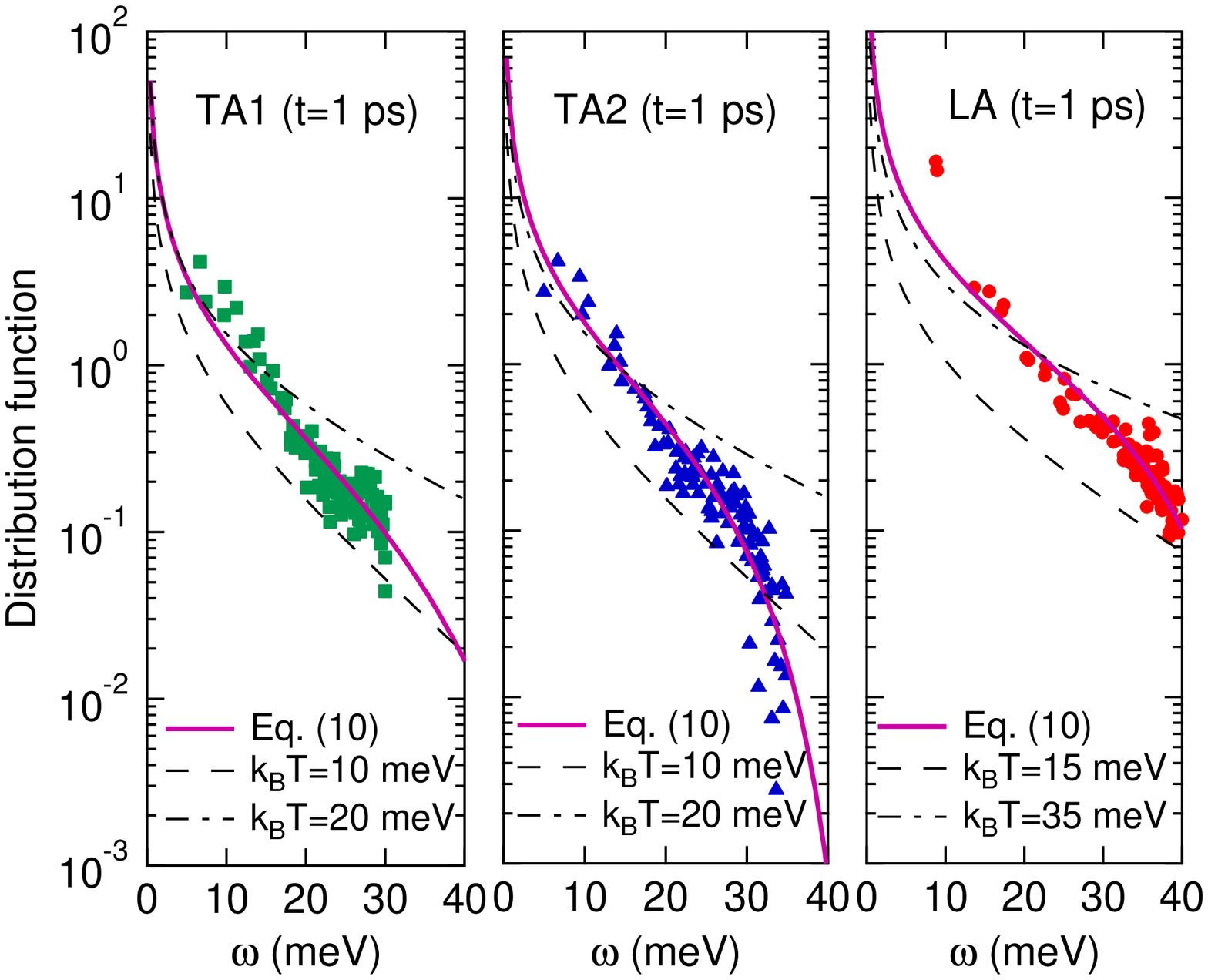}
\caption{\label{fig:1ps_app} Same as Fig.~\ref{fig:1ps} but for the case of $f=w_{0}^{2}$.}
\end{figure}
%%%%%%%%%%%%%%%%%
%%%%%%%%%%%%%%%%%
\begin{figure}[ttt]
\center
\includegraphics[scale=0.4,clip]{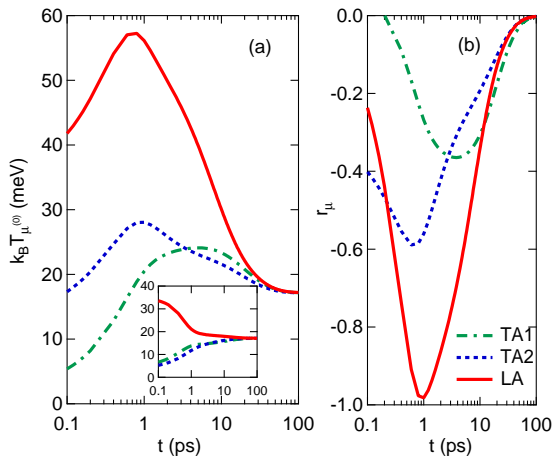}
\caption{\label{fig:kTr_app} Same as Fig.~\ref{fig:kTr} but for the case of $f=w_{0}^{2}$.}
\end{figure}
%%%%%%%%%%%%%%%%%
%%%%%%%%%%%%%%%%%
\begin{figure}[ttt]
\center
\includegraphics[scale=0.45,clip]{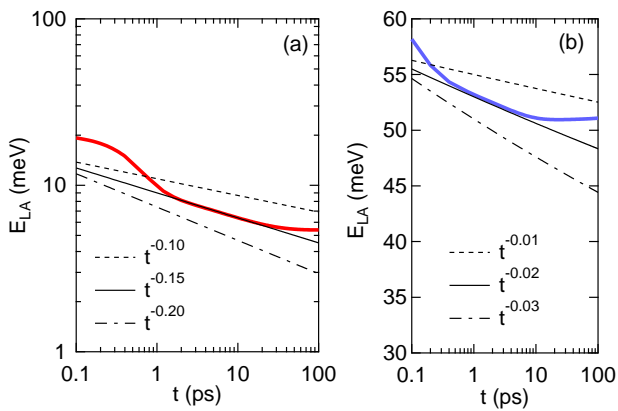}
\caption{\label{fig:power_app} The time-evolution of the LA phonon energy $E_{\rm LA}(t)$ for the initial conditions; $(k_{\rm B}T_{\rm low}, k_{\rm B}T_{\rm high})=$(1,35) meV for (a) and (60,80) meV for (b). The curves proportional to $t^{-\alpha}$ with $\alpha=$0.10, 0.15, and 0.20 for (a) and 0.01, 0.02, and 0.03 for (b) are also shown. $f=w_{0}^{2}$ is used.}
\end{figure}
%%%%%%%%%%%%%%%%%

%%%%%%%%%%%%%%%%%%%%%%%%%%%%%%%%%%%%
\section{Nonlinear least square problem}
%%%%%%%%%%%%%%%%%%%%%%%%%%%%%%%%%%%%
\label{app:nonlinear}
The numerical solution of the BTE in Eq.~(\ref{eq:BTE_ph}) gives the phonon occupation numbers $n_{\bm{q},\mu}(t)$ at the phonon energy $\hbar\omega_{\bm{q},\mu}$. At each $t$, we minimize the following function with respect to $T_{\mu}^{(0)}$ and $r_\mu$,
\begin{eqnarray}
 & &G(T_{\mu}^{(0)},r_\mu,t) 
 \nonumber\\
 &=& \sum_{\bm{q}} \left[ \ln \frac{n_{\bm{q},\mu}(t)}
 {
 n^{(0)}(\omega_{\bm{q},\mu}, T_{\mu}^{(0)} + r_{\mu} \frac{\hbar \omega_{\bm{q},\mu}}{k_{\rm B}})
 }\right]^2,
 \label{eq:GT}
\end{eqnarray}
where $n^{(0)}(\omega,T) = \left[ \exp(\hbar\omega/k_{\rm B}T) -1 \right]^{-1}$ is the Bose-Einstein distribution function. $n^{(0)}(\omega_{\bm{q},\mu}, T_{\mu}^{(0)} + r_{\mu} \frac{\hbar \omega_{\bm{q},\mu}}{k_{\rm B}})$ in Eq.~(\ref{eq:GT}) is equivalent to Eq.~(\ref{eq:FDT}). This minimization problem is equivalent to the nonlinear least square problem and can be solved by using \texttt{minpack} \cite{minpack}. The logarithm in bracket in Eq.~(\ref{eq:GT}) should be taken in order to lower the magnitude of errors at each $\hbar\omega_{\bm{q},\mu}$. The solid curves in Figs.~\ref{fig:1ps} and \ref{fig:1ps_app} (below) and the all curves drawn in Figs.~\ref{fig:kTr} and \ref{fig:kTr_app} (below) were obtained by employing this method. 

%%%%%%%%%%%%%%%%%%%%%%%%%%%%%%%%%%%%
\section{Numerical results for the Boltzmann transport equation: A case of $f=w_{0}^{2}$}
%%%%%%%%%%%%%%%%%%%%%%%%%%%%%%%%%%%%
\label{sec:fconst}
By assuming the constant coupling function $f=w_{0}^{2}$, that is, using $\vert M_{\bm{q},\bm{q}_1,\bm{q}_2}^{\mu,\mu_1,\mu_2} \vert^2 = \delta_{\Delta \bm{q},\bm{G}} w_{0}^{2}$ with $w_0 = 2$ meV, the BTE of Eq.~(\ref{eq:BTE_ph}) is solved numerically. All the other parameters are the same in the main text. Corresponding to the main text, Figure \ref{fig:time-evolution_app} shows the distribution of TA1, TA2, and LA modes for $t=0.1, 1, 10$, and 100 ps; Figure \ref{fig:1ps_app} shows the phonon occupation numbers at $t=$1 ps; Figure \ref{fig:kTr_app} shows the $t$-dependence of $T_{\mu}^{(0)}$ and $r_{\mu}$; Figures \ref{fig:power_app}(a) and \ref{fig:power_app}(b) show the time-evolution of the total LA phonon energy per a unit cell for the initial conditions $(k_{\rm B}T_{\rm low}, k_{\rm B}T_{\rm high})=$(1,35) meV and (60,80) meV, respectively. The small value of the exponent $p=0.02$ is observed for the latter case because the ratio of $E_{\rm LA}(t = 0.1 {\rm ps})/E_{\rm LA}(t = 100 {\rm ps})$ is relatively small, compared to the former case. Contrary to the case of $f= w_{0}^2 \vert \bm{Q} \vert \vert \bm{Q}_1 \vert \vert \bm{Q}_2 \vert$, that is, Eq.~(\ref{eq:matrix}), the suppression of the population of the low-energy TA phonons is not observed, as shown in Figs.~\ref{fig:time-evolution_app} and \ref{fig:1ps_app}. Apart from this, overall features in Figs.~\ref{fig:time-evolution_app}, \ref{fig:1ps_app}, \ref{fig:kTr_app}, and \ref{fig:power_app} are almost the same as Figs.~\ref{fig:time-evolution}, \ref{fig:1ps}, \ref{fig:kTr}, and \ref{fig:power}, respectively, irrespective to the different form of the matrix elements. This may imply that the details of the matrix elements ({\it i.e.}, $\bm{q}$-dependence) do not play a major role in the phonon thermalization, while the systematic investigations with the use of more realistic $f$ are desired.

%===================================================================%
%   References
%===================================================================%


\begin{thebibliography}{99}

\bibitem{anisimov} S. I. Anisimov, B. L. Kapeliovich, and T. L. Perel'man, Electron emission from the metal surfaces induced by ultrashort lasers pulses, Zh. Eksp. Teor. Fiz. {\bf 66}, 776 (1974) [Sov. Phys. JETP {\bf 39}, 375 (1974)].

\bibitem{allen} P. B. Allen, Theory of thermal relaxation of electrons in metals, Phys. Rev. Lett. {\bf 59}, 1460 (1987).

\bibitem{brorson} S. D. Brorson, A. Kazeroonian, J. S. Moodera, D. W. Face, T. K. Cheng, E. P. Ippen, M. S. Dresselhaus, and G. Dresselhaus, Femtosecond room-temperature measurement of the electron-phonon coupling constant $\mu$ in metallic superconductors, Phys. Rev. Lett. {\bf 64}, 2172 (1990).

\bibitem{lin2008} Z. Lin, L. V. Zhigilei, and V. Celli, Electron-phonon coupling and electron heat capacity of metals under conditions of strong electron-phonon nonequilibrium, Phys. Rev. B {\bf 77}, 075133 (2008).

\bibitem{brown2016} A. M. Brown, R. Sundararaman, P. Narang, W. A. Goddard, III, and H. A. Atwater, Ab initio phonon coupling and optical response of hot electrons in plasmonic metals, Phys. Rev. B {\bf 94}, 075120 (2016).

\bibitem{nakamura} A. Nakamura, T. Shimojima, M. Nakano, Y. Iwasa, and K. Ishizaka, Electron and lattice dynamics of transition metal thin films observed by ultrafast electron diffraction and transient optical measurements, Struct. Dyn. {\bf 3}, 064501 (2016).

\bibitem{bistritzer} R. Bistritzer and A. H. MacDonald, Electronic Cooling in Graphene, Phys. Rev. Lett. {\bf 102}, 206410 (2009).

\bibitem{vilijas} J. K. Viljas and T. T. Heikkil\"{a}, Electron-phonon heat transfer in monolayer and bilayer graphene, Phys. Rev. B {\bf 81}, 245404 (2010).

\bibitem{ono3} S. Ono, Y. Toda, and J. Onoe, Unified understanding of the electron-phonon coupling strength for nanocarbon allotropes, Phys. Rev. B {\bf 90}, 155435 (2014).

\bibitem{lundgren} R. Lundgren and G. A. Fiete, Electronic cooling in Weyl and Dirac semimetals, Phys. Rev. B {\bf 92}, 125139 (2015).

\bibitem{white2014} T. G. White, N. J. Hartley, B. Borm, B. J. B. Crowley, J. W. O. Harris, D. C. Hochhaus, T. Kaempfer, K. Li, P. Neumayer, L. K. Pattison, F. Pfeifer, S. Richardson, A. P. L. Robinson, I. Uschmann, and G. Gregori, Electron-Ion Equilibration in Ultrafast Heated Graphite, Phys. Rev. Lett. {\bf 112}, 145005 (2014).

\bibitem{groeneveld} R. H. M. Groeneveld, R. Sprik, and Ad. Lagendijk, Femtosecond spectroscopy of electron-electron and electron-phonon energy relaxation in Ag and Au, Phys. Rev. B {\bf 51}, 11433 (1995).

\bibitem{rethfeld} B. Rethfeld, A. Kaiser, M. Vicanek, and G. Simon, Ultrafast dynamics of  nonequilibrium electrons in metals under femtosecond laser irradiation, Phys. Rev. B {\bf 65}, 214303 (2002).

\bibitem{kabanov2008} V. V. Kabanov and A. S. Alexandrov, Electron relaxation in metals: Theory and exact analytical solutions, Phys. Rev. B {\bf 78}, 174514 (2008).

\bibitem{ishida2011} Y. Ishida, T. Togashi, K. Yamamoto, M. Tanaka, T. Taniuchi, T. Kiss, M. Nakajima, T. Suemoto, and S. Shin, Non-thermal hot electrons ultrafastly generating hot optical phonons in graphite, Sci. Rep. {\bf 1}, 64 (2011).

\bibitem{mueller} B. Y. Mueller and B. Rethfeld, Relaxation dynamics in laser-excited metals under nonequilibrium conditions, Phys. Rev. B {\bf 87}, 035139 (2013).

\bibitem{baranov} V. V. Baranov and V. V. Kabanov, Theory of electronic relaxation in a metal excited by an ultrafast optical pump, Phys. Rev. B {\bf 89}, 125102 (2014).

\bibitem{waldecker} L. Waldecker, R. Bertoni, and R. Ernstorfer, and J. Vorberger, Electron-Phonon Coupling and Energy Flow in a Simple Metal beyond the Two-Temperature Approximation, Phys. Rev. X {\bf 6}, 021003 (2016).

\bibitem{ishida2016} Y. Ishida, H. Masuda, H. Sakai, S. Ishiwata, and S. Shin, Revealing the ultrafast light-to-matter energy conversion before heat diffusion in a layered Dirac semimetal, Phys. Rev. B {\bf 93}, 100302(R) (2016).

\bibitem{trigo} M. Trigo, J. Chen, V. H. Vishwanath, Y. M. Sheu, T. Graber, R. Henning, and D. A. Reis, Imaging nonequilibrium atomic vibrations with x-ray diffuse scattering, Phys. Rev. B {\bf 82}, 235205 (2010).

\bibitem{harb} M. Harb, H. Enquist, A. Jurgilaitis, F. T. Tuyakova, A. N. Obraztsov, and J. Larsson, Phonon-phonon interactions in photoexcited graphite studied by ultrafast electron diffraction, Phys. Rev. B {\bf 93}, 104104 (2016).

\bibitem{shin} S. Shin and M. Kaviany, Optical phonon production by upconversion: Heterojunction-transmitted versus native phonons, Phys. Rev. B {\bf 91}, 165310 (2015).

\bibitem{fahy} S. Fahy, \'{E}. D. Murray, and D. A. Reis, Resonant squeezing and the anharmonic decay of coherent phonons, Phys. Rev. B {\bf 93}, 134308 (2016).

%\bibitem{cepellotti} A. Cepellotti and N. Marzari, Thermal Transport in Crystals as a Kinetic Theory of Relaxons, Phys. Rev. X {\bf 6} 041013 (2016).

%\bibitem{tamura} S. Tamura, Spontaneous decay rates of LA phonons in quasi-isotropic solids, Phys. Rev. B {\bf 31}, 2574(R) (1985).

%\bibitem{narasimhan} S. Narasimhan and D. Vanderbilt, Anharmonic self-energies of phonons in silicon, Phys. Rev. B {\bf 43}, 4541(R) (1990).

%\bibitem{debernardi} A. Debernardi, S. Baroni, and E. Molinari, Anharmonic Phonon Lifetimes in Semiconductors from Density-Functional Perturbation Theory, Phys. Rev. Lett. {\bf 75}, 1819 (1995).

\bibitem{ashcroft_mermin} N. W. Ashcroft, N. D. Mermin, and D. Wei, {\it Solid State Physics}, revised edition, Cengage Learning (2016).

%\bibitem{suppl} See Supplemental Material at XXX for the phonon model used in the numerical simulation, the determination of two parameters given in Eq.~(\ref{eq:FDT}), the numerical results in the case of $f = w_{0}^{2}$, the evaluation of the collision term at the initial stage, and the derivation and analytical solution of Eq.~(\ref{eq:theta}).

\bibitem{ziman} J. M. Ziman, {\it Electrons and Phonons} (Oxford University Press, 1960).

\bibitem{landau} L. D. Landau and E. M. Lifshitz, {\it Physical Kinetics} (Pergamon Press, 1981).

\bibitem{bonini} N. Bonini, M. Lazzeri, N. Marzari, and F. Mauri, Phonon Anharmonicities in Graphite and Graphene, Phys. Rev. Lett. {\bf 99}, 176802 (2007).

\bibitem{tersoff} J. Tersoff, New empirical approach for the structure and energy of covalent systems, Phys. Rev. B {\bf 37}, 6991 (1988).

\bibitem{MP} H. J. Monkhorst and J. D. Pack, Special points for Brillouin-zone integrations, Phys. Rev. B {\bf 13}, 5188 (1976).

\bibitem{minpack} http://www.netlib.org/minpack/

\bibitem{SCs} Similar model has been used in studying the nonequilibrium phonons of superconductors and charge-density-wave materials \cite{kabanov1999,ono2012}.

\bibitem{kabanov1999} V. V. Kabanov, J. Demsar, B. Podobnik, and D. Mihailovic, Quasiparticle relaxation dynamics in superconductors with different gap structures: Theory and experiments on YBa$_2$Cu$_3$O$_{7-\delta}$, Phys. Rev. B {\bf 59}, 1497 (1999).

\bibitem{ono2012} S. Ono, H. Shima, and Y. Toda, Theory of photoexcited carrier relaxation across the energy gap of phase-ordered materials, Phys. Rev. B {\bf 86}, 104512 (2012).

\bibitem{rothwarf} A. Rothwarf and B. N. Taylor, Measurement of Recombination Lifetimes in Superconductors, Phys. Rev. Lett. {\bf 19}, 27 (1967).

%\bibitem{maradudin} A. A. Maradudin, {\it Dynamical Properties of Solids}, edited by G. K. Horton and A. A. Maradudin (North-Holland, Amsterdam, 1974).

%\bibitem{michel} K. H. Michel, S. Costamagna, and F. M. Peeters, Theory of anharmonic phonons in two-dimensional crystals, Phys. Rev. B {\bf 91}, 134302 (2015).

\end{thebibliography}
\end{document}